# Tetrathiafulvalene radical cation (TTF[+·]) Charge Transfer aggregates included in PMMA matrix, a Resonance Raman Study

B. F. Scremin[1]

1-IOM CNR, Institute "Officina dei Materiali" of the National Research Council of Italy, Tasc Laboratory, Strada Statale 14, Km 163.5, Trieste Italy

barbara.scremin@cnr.it, scremin@iom.cnr.it

**Abstract.**

*TTFClO$_4$ salts were included in a PMMA matrix via solvent casting technique under vacuum, up to 72 hours of pumping. The synthesis of the salt, present in the literature was enriched with a further purification procedure, wich resulted important to obtain TTF[+·] radical cation aggregates. The thin films were characterized by low temperature UV-VIS-NIR absorption, and a multicomponent broad charge transfer band (CT) in the near infrared was found. Raman spectroscopy (RRS) in resonance with the CT transition allowed to study the origin of the CT absorption: the species present in the films were recognized as dimers and aggregates of TTF[+·] radical cation.*

Tetrathiafulvalene radical cation (TTF[+·]) salts [4] in solution with polar solvents (dimethyl sulfoxide DMSO [1], dimethyl formammide DMF [5]) form radical dimers and aggregated systems showing a broad charge transfer (CT) absorption band around 800 nm.

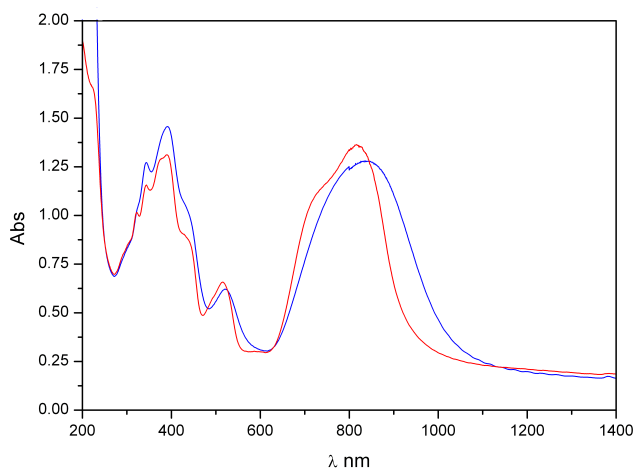

Fig. 1 Absorption spectra of TTFClO$_4$ included in PMMA matrix: in red the low temperature spectrum, in blue room temperature spectrum.

Therefore, the stability of solutions is limited in time. The mechanism of TTFClO$_4$ degradation was investigated and the preparation protocol improved on this basis. Solutions of the salt were prepared and it was noticed that the strong CT band disappeared with time; the process was monitored by UV-Vis-NIR absorption and it was noticed that the final spectra corresponded to that of TTF$^0$ (there is also a hypothesis of oxygen contribution and the formation of an S-Oxyde [3]), the molecule from which the TTFClO$_4$ was synthesized.

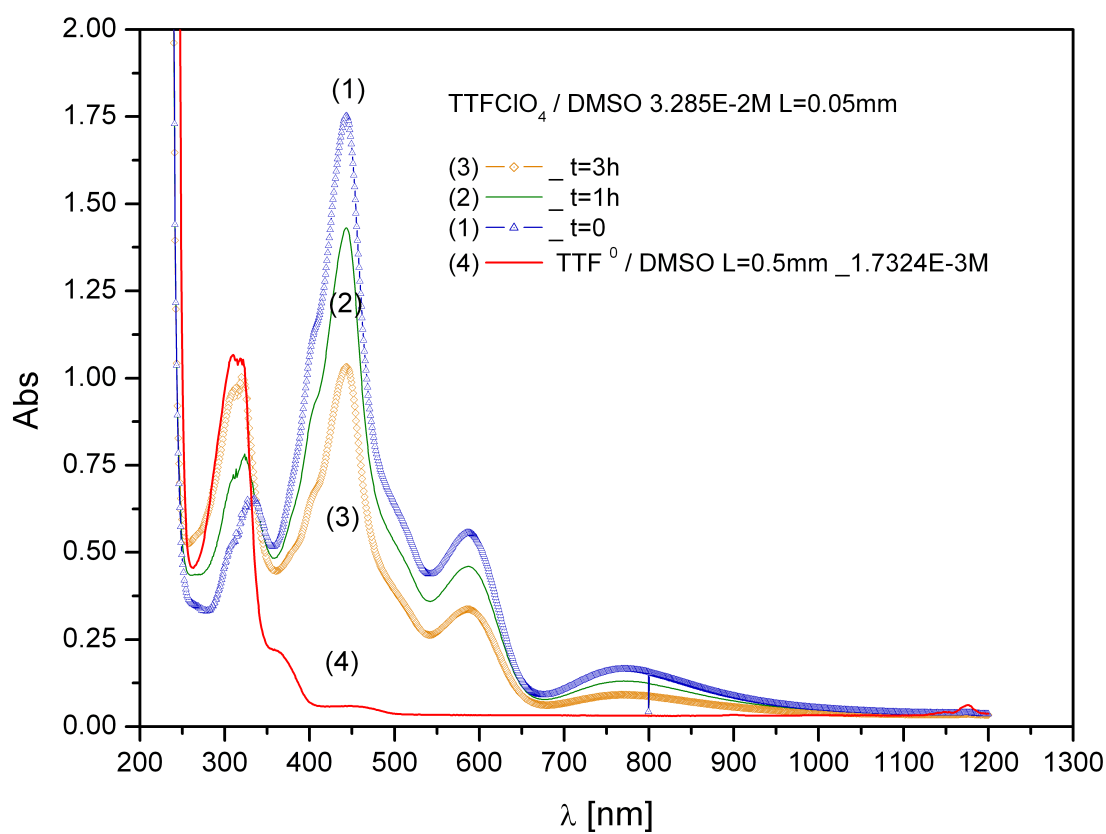

Fig. 2 Spectra of TTFClO$_4$ in DMSO solutions as function of time. For reference TTF$^0$ spectrum (red).

The chemical reaction was recognized as a dismutation of TTF$^+$ in TTF$^0$ and TTF$^{2+}$ the latter giving rise to highly insoluble salt that subtract the cation from the solution equilibria [3]. In fact, given this information the TTFClO$_4$ salt was prepared according to the literature method [2] and a purification procedure was proposed. TTFClO$_4$ synthesized was washed about three times with cyclohexane solubilizing TTF$^0$ and filtered with gooch G4 or G5, which had to be washed with a concentrated solution of NaOH to remove TTF$^{2+}$ salt from the filter pores. Finally, the solution of the TTFClO$_4$ in CH$_3$CN was dried under vacuum, since it is supposed that TTFClO$_4$ is poorly stable under atmosphere [4, 5]. The purification procedure on the basis of RRS spectra of doped PMMA film suggested that it was important to remove TTF(ClO$_4$)$_2$ which gives rise to nucleation centres for the crystallization of TTFClO$_4$ in the PMMA matrix, instead of forming TTF$^+$ dimers and higher order aggregates. This was evidenced by RRS spectroscopy in resonance with the CT transition [5]. The doped PMMA films were prepared in N$_2$ atmosphere using a solution about 0.2 M of TTFClO$_4$ in DMSO and a PMMA solution in CH$_3$Cl about 275 g/l. The less viscous TTFClO$_4$ solution was transferred in PMMA one in a volumetric ratio 3.75:1 to obtain a final 0.27 M TTF$^+$ concentration in PMMA. Finally, 65-80 µl of the final solution was deposited on a clean (with piranha solution) quartz window inside a vial connected to a vacuum line (about 10$^{-6}$ mbar) equipped with liquid nitrogen trap [4]. The solvents evaporation had to be gradual to avoid their boiling, and the drying process was carried out up to 72 hours.

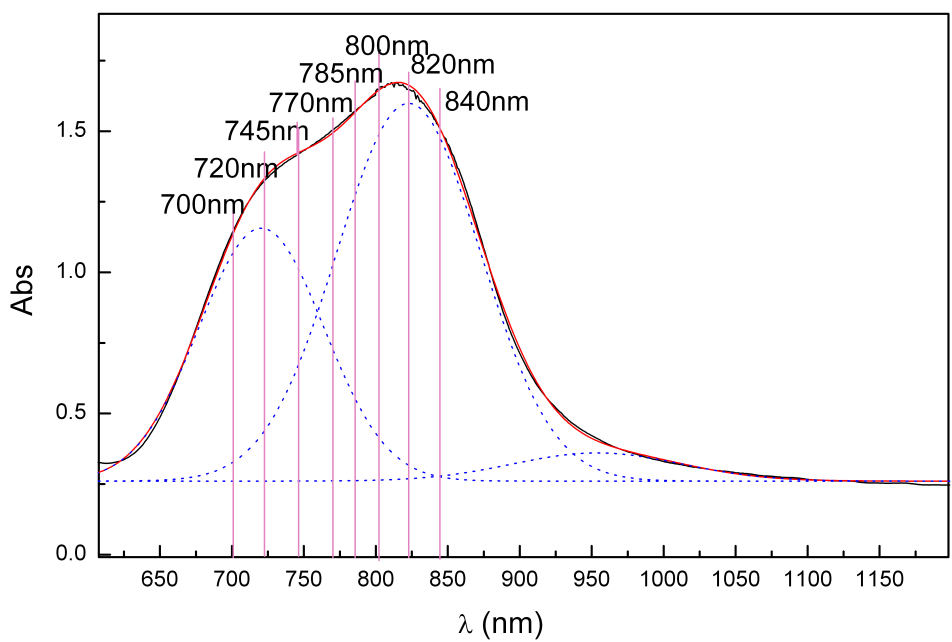

Fig. 3 The TTFClO$_4$/PMMA CT band at 15K (nominal value), the laser lines from Ti:Sapphire laser used to excite the RRS spectra (intermolecular phonons modes of TTF$^{+.}$) are shown for reference.

The final thickness of the films was about 20-30 µm (measured with a micrometer). The technique was not suitable for preparation of thicker samples. Solvents like CH3CN were tested for TTFClO$_4$ but they resulted the precipitation of the salt in PMMA matrix [5]. DMF gave rise to poor optical quality films [5]. PMMA films doped with TTFClO$_4$ presented a broad (and complex, Fig. 3) and strong absorption band around 800 nm Fig. 3. The film absorption spectrum was acquired (Cary 5, Varian) at low temperature (15 K) and under vacuum with a cryostat (CTI Cryogenics-Cryodine SC) and was fitted with three Gaussians: the first component was at 720 nm and the second at 822 nm and the weak third around 970 nm. In Fig. 3 the CT band is shown together with the wavelengths used for the excitation of RRS spectra of the phonon modes of TTF$^{+\cdot}$ acquired to investigate the nature of the double (triple) CT absorption.

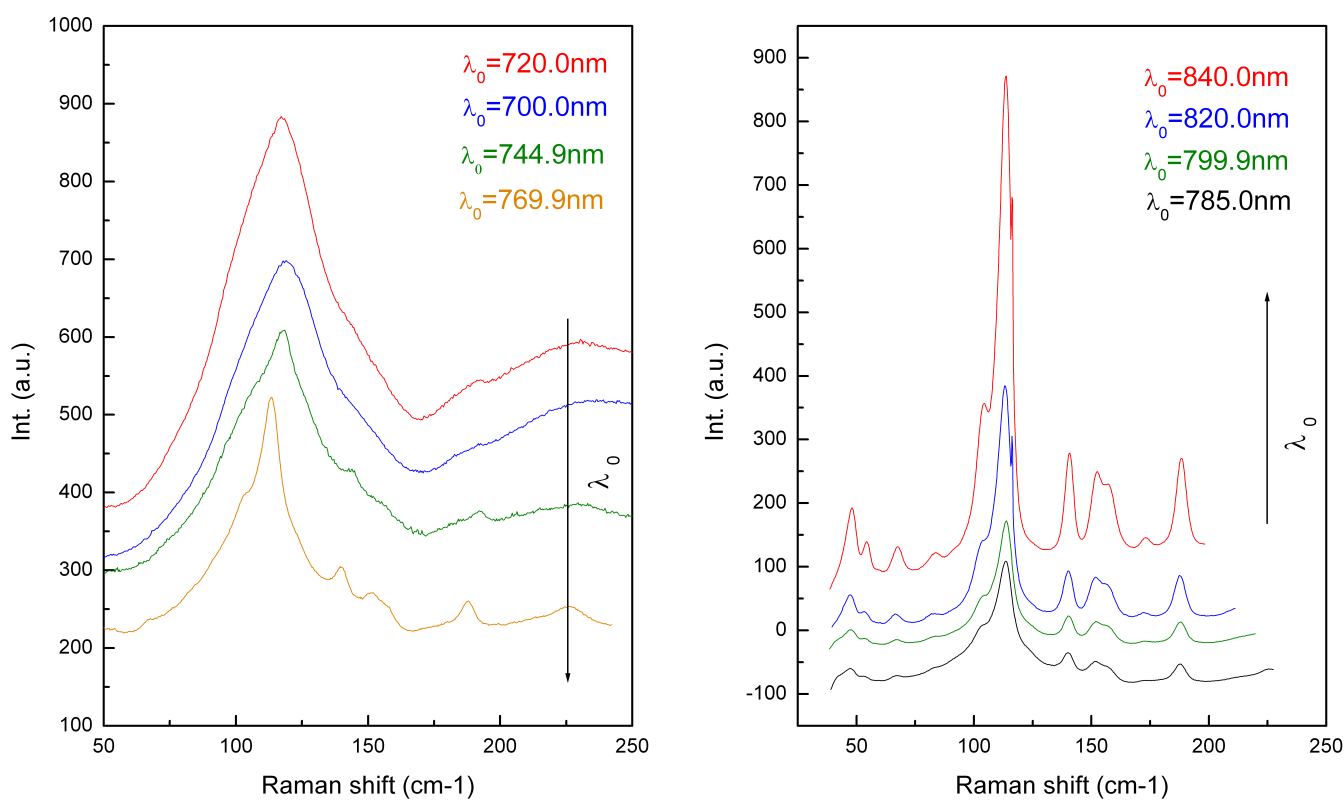

Fig. 4 RRS spectra for TTFClO$_4$/PMMA system, phonon modes of TTF$^{+\cdot}$, left- dimers, right- higher order aggregates.

The Resonant Raman Spectroscopy system consisted of an Argon gas laser (Innova 310, Coherent) pumping a CW Titanium: Sapphire (3900S Coherent) solid state laser with tuneable emission around 800 nm. A triple monochromator (S 3000 Jobin Yvon) equipped with liquid nitrogen cooled CCD camera analysed the Raman scattered radiation; the experiment on films was performed at low temperature (15K) with the use of a cryostat. The results are shown in Fig. 4 The first set of spectra in Fig.3 was resonant with the first CT component and comparing with RRS spectra of solutions of TTFClO$_4$ in DMSO (Fig. 5) and then was associated with the presence of TTF$^{+\cdot}$ dimers, while the lower energy CT component was assigned to higher order aggregates, peculiar aspect of the system formed in doped PMMA films dried for 72 hours. In Fig. 5 are shown the spectra of TTFClO$_4$ crystalline powders [4] suspended in a nujol mull (at 15K) and crystalline powders dissolved in DMSO -substantially representing the crystal [6] RRS and the dimers [4] respectively; these RRS spectra were acquired for comparison with the RRS of the films: it allowed to assign to the dimers the higher energy component in the CT band and to aggregates, and not to the hypothetically crystallized salt, the lower energy component.

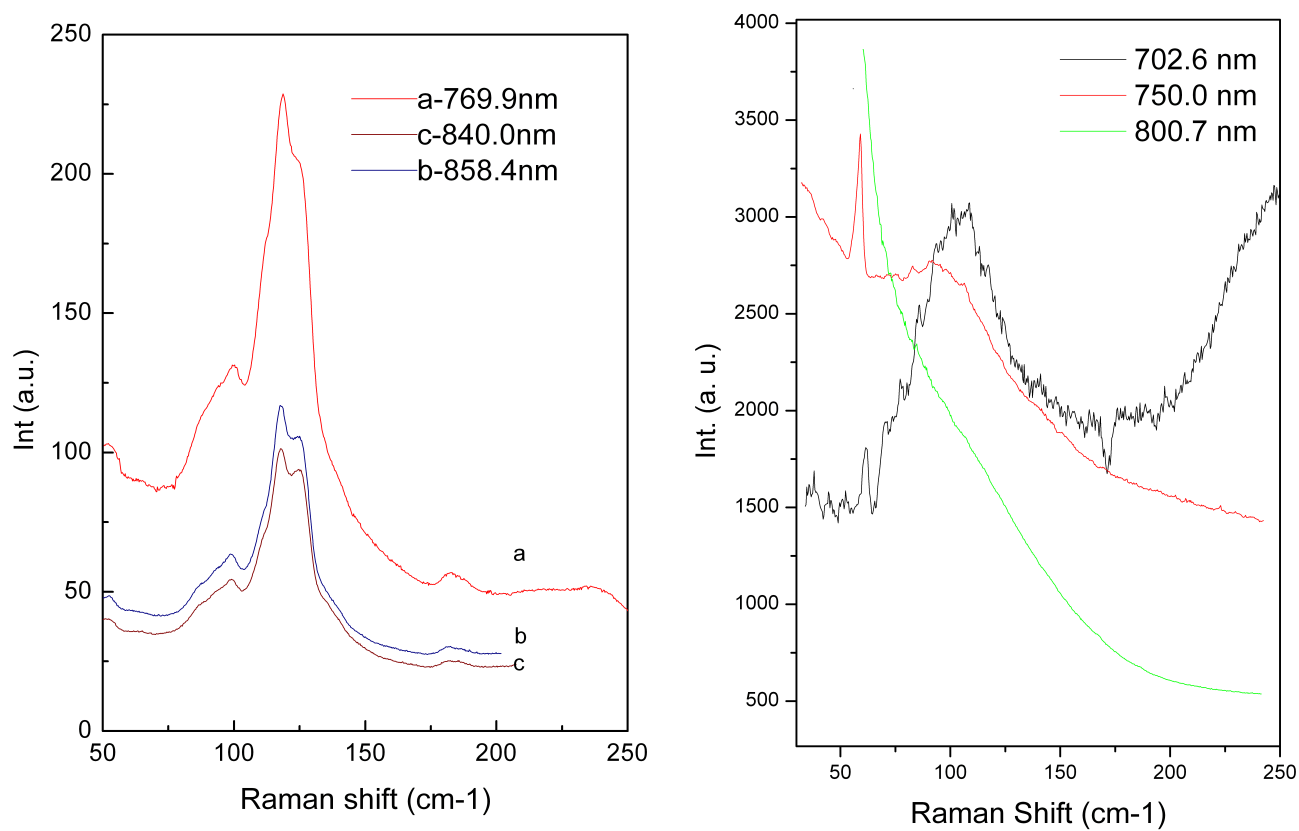

Fig. 5 RRS spectra of crystalline powders TTFClO$_4$ at 15 K (left) and of a 0.2 M TTFClO$_4$ solution in DMSO (right)- Varian Cary 5 [4].

Molar extinctions coefficients for TTF$^0$ and the synthesized TTF S-Oxide [6] are reported in Fig. 6 since under air exposure both species can be formed from TTF$^{+\cdot}$ as can be seen from Fig. 2 and explained in ref. [3], where absorption spectra are reported for film exposed to air up to an equilibrium. Here all the RRS spectra were acquired with the film under vacuum.

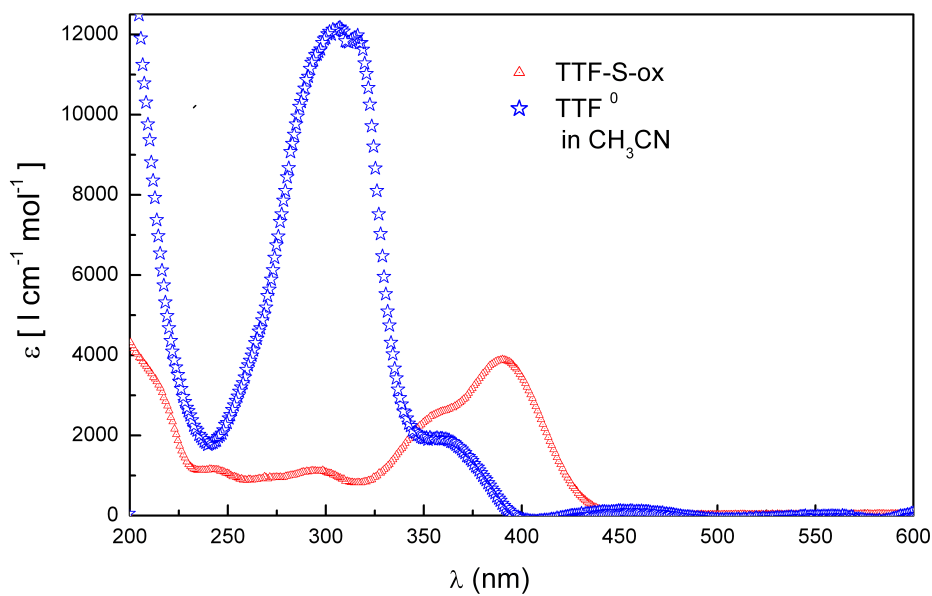

Fig. 6. Molar extinctions coefficients for TTF$^0$ and TTF S-Oxide in CH$_3$CN [5], Varian Cary 5, ambient temperature.

It was observed for TTFClO$_4$ synthesized with the literature procedure [2] without the purification steps here described, that RRS spectra of films in particular the ones in resonance with the lower energy component showed, together with the phononic pattern in Fig. 4 (right) of aggregates also the peaks found for the crystalline powders [7, 8] (Fig. 5 left side) particularly evident for the more intense peaks [5].

**Conclusions**

TTFClO$_4$ /PMMA film system was prepared and studied. It was found that it was possible to obtain dimers of TTF$^{+\cdot}$ and higher order CT aggregates. UV-VIS absorption spectroscopy at low temperature evidenced three components in the CT absorption band in the near infrared, and with RRS performed in resonance with the two principal components of the CT band allowed to study the phonon modes of TTF$^{+\cdot}$ in PMMA. Comparing RRS phonon spectra of solutions and of crystalline powders it was possible to assign the higher energy component of the CT band to dimers (comparing with solutions) and the lower energy component to the presence of higher order aggregates (comparing with solutions and crystalline powders RRS spectra). The study of the weak third component of the CT band was not experimentally feasible. The result is of interest since it was the first time that such a system was prepared and demonstrated, this is a reason why few literature references are reported.

**Conflict of interests**

BF Scremin declares no conflict of interests

**Acknowledgments**

Financial support was from Padova University, Physical Chemistry Department, BFS acknowledge the financial support of a PhD fellowship of the Padova University. Advises and discussions with colleagues, Mrs. Miranda Zanetti of the department are acknowledged. The supervision of Prof. Renato Bozio was greatly acknowledged.

**Notes and references**